\documentclass[a4paper]{ceurart}

\usepackage{amsmath,amssymb,amsfonts}
\usepackage{algorithmic}
\usepackage{graphicx}
\usepackage{textcomp}
\usepackage{xcolor}

\usepackage{tikz}
\usepackage[capitalize]{cleveref}
\usetikzlibrary{shapes,arrows}
\usepackage{booktabs}
\usepackage{siunitx}
\usepackage{caption}
\usepackage{subcaption}
\usepackage{orcidlink}
\usepackage{hyperref}

\begin{document}

\copyrightyear{2025}
\copyrightclause{Copyright for this paper by its authors.
  Use permitted under Creative Commons License Attribution 4.0
  International (CC BY 4.0).}

\conference{IPIN-WCAL 2025: Workshop for Computing \& Advanced Localization at the Fifteenth International Conference on Indoor Positioning and Indoor Navigation, September 15--18, 2025, Tampere, Finland}

\title{Range-Angle Likelihood Maps for Indoor Positioning Using Deep Neural Networks}

\author[1]{Muhammad Ammad}[%
orcid=0009-0007-2720-8433,
email=Muhammad.Ammad@tu-dresden.de,
]
\author[1]{Paul Schwarzbach}[%
orcid=0000-0002-1091-782X,
]
\author[2]{Michael Schultz}[%
orcid=0000-0003-4056-8461,
]
\author[1]{Oliver Michler}[%
orcid=0000-0002-8599-5304,
]

\address[1]{Chair of Transport Systems Information Technology, Institute of Traffic Telematics, Technische Universit\"at Dresden}
\address[2]{Chair of Air Traffic Concepts, University of the Bundeswehr Munich}

\begin{abstract}
Accurate and high precision of the indoor positioning is as important as ensuring reliable navigation in outdoor environments. Using the state-of-the-art deep learning models provides better reliability and accuracy to navigate and monitor the accurate positions in the aircraft cabin environment. We utilize the simulated aircraft cabin environment measurements and propose a residual neural network (ResNet) model to predict the accurate positions inside the cabin. The measurements include the ranges and angles between a tag and the anchors points which are then mapped onto a grid as range and angle residuals. These residual maps are then transformed into the likelihood grid maps where each cell of the grid shows the likelihood of being a true location. These grid maps along with the true positions are then passed as inputs to train the ResNet model. Since any deep learning model involve numerous parameter settings, hyperparameter optimization is performed to get the optimal parameters for training the model effectively with the highest accuracy. Once we get the best hyperparameters settings of the model, it is then trained to predict the positions which provides a centimeter-level accuracy of the localization.
\end{abstract}


\maketitle

\vspace{-0.5cm}
\section{Introduction}
\label{sec:Intro}

Indoor localization systems (IPS) have become increasingly important in recent advancements, particularly due to the major challenge of limited availability of satellite positioning systems in indoor environments \cite{Yassin2017}. Technologies such as Wi-Fi, Ultra-Wideband (UWB), Bluetooth Low Energy (BLE), and ZigBee have gained attention as viable solutions for precise indoor positioning \cite{Hsieh2019}\cite{Niu2023}. Recently, the integration of these systems with the artificial intelligence (AI) has gathered significant attention, primarily due to the increased precision and higher accuracy of their positioning outcomes \cite{Nessa2020}.

The application of deterministic radio propagation simulations to improve AI-enabled localization and sensing systems is explored by identifying existing bottlenecks in the conventional tool-chain and suggesting advancements in simulation techniques \cite{Michler2023PotentialsRadioSimulation}. To optimize the aircraft in-cabin operations and improving passenger experience, simulation techniques has been employed to assess positioning accuracy and system reliability \cite{Schwarzbach2024SimulationIPSConnectedCabin}. The current study further investigates the simulation-based evaluation of IPS tailored for connected aircraft cabins using state-of-the-art deep learning techniques.

\begin{figure}[pos=ht]
    \centering
    \includegraphics[width=0.8\linewidth]{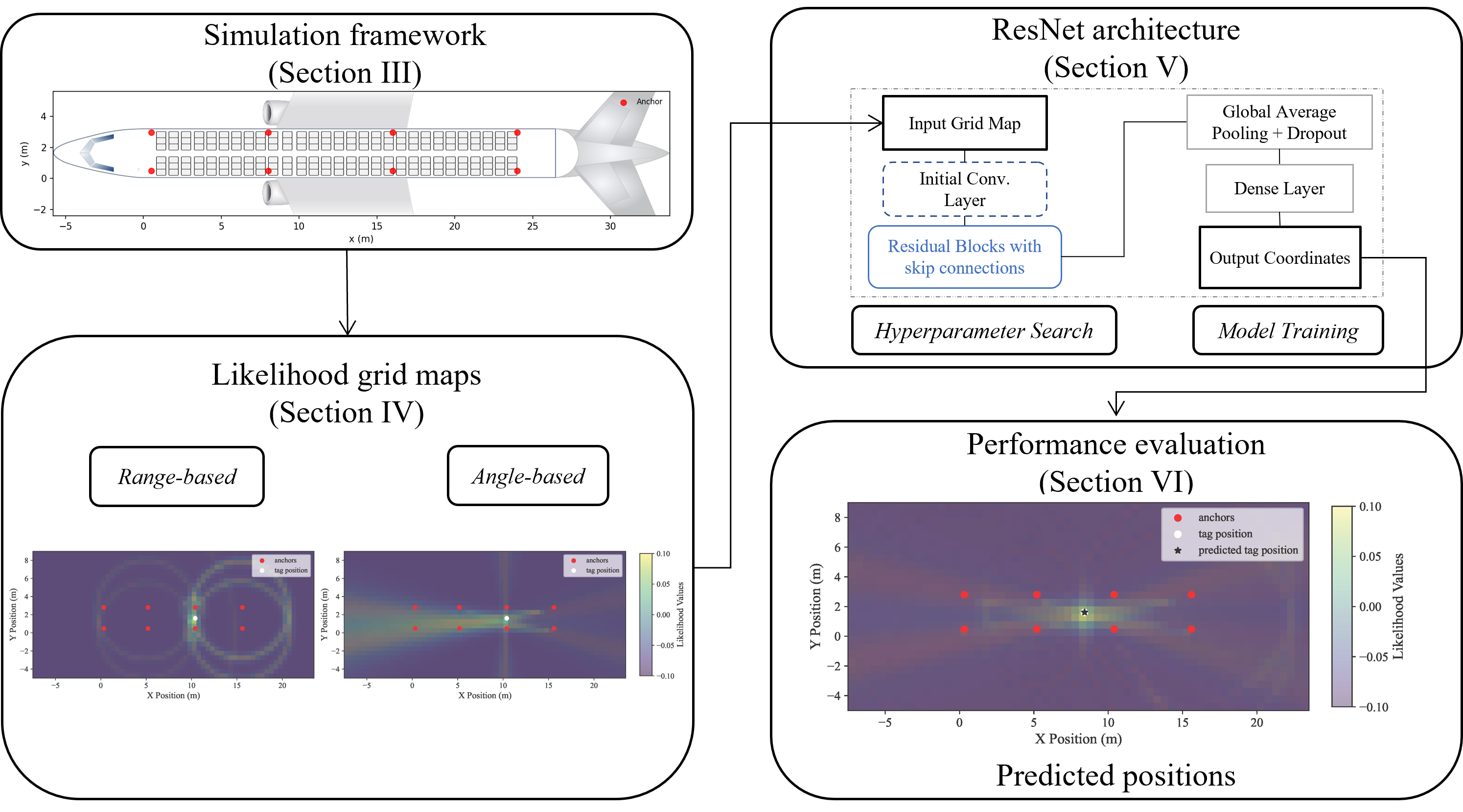}
    \caption{The illustrative diagram representing a comprehensive overview of the research plan to predict location coordinates based on the range and angle likelihood grid maps. These grid maps are the backbone of this research for localization using ResNet model shown on the right side.}
    \label{fig:concept_diagram}
\end{figure}

As illustrated in \cref{fig:concept_diagram}, in this paper, we use the data-driven localization methods utilizing the simulated range and angle of arrival (AoA) measurements based on observability from different anchor points. These measures are then used to construct the likelihood grid maps using the residuals, which serve as spatial representations of the passenger's possible location. Then the ResNet \cite{resnet_ref} model is trained on these grid maps to analyze the spatial data thus predicting the passenger's location.

The remainder of the paper is structured as follows: \cref{sec:related_work} reviews related work and background of different technologies. \cref{sec:data_simulation} describes the data simulation methodology, aircraft environment setup, and error modeling. \cref{sec:grid_maps} shows how the range and angle residuals are calculated to generate the likelihood grid maps. \cref{sec:positioning} explores in-depth details of the ResNet architecture layers along with the input and loss function. \cref{sec:evaluation} discusses the hyperparameter tuning impacts on the accuracy and finally provides a comprehensive evaluation of the model's accuracy on the dataset.
\section{Related Work}
\label{sec:related_work}
The application of AI and machine learning (ML) technologies in indoor positioning has attracted considerable interest in recent research due to their ability to enhance accuracy, reliability and accuracy \cite{Pei2024}. Indoor positioning methods can be categorized into two key paradigms: (1) direct AI/ML positioning, and (2) AI/ML-assisted positioning \cite{yang2024positioningusingwirelessnetworks}. The first positioning method estimates the location using raw sensor data like Wi-Fi and Bluetooth signals, employing ML algorithms such as neural networks, decision trees, and support vector machines (SVM) to detect complex patterns without traditional signal processing \cite{Bozkurt2015}. This approach streamlines localization but relies on extensive training data to avoid overfitting issues which may arise during training process \cite{Nessa2020}, but obtaining a robust training dataset remains a challenge.

In contrast, AI/ML-assisted positioning enhances traditional methods for localization by integrating ML for error correction and adaptive learning. This hybrid approach strengthens the reliability of conventional triangulation or fingerprint-based localization \cite{yang2024positioningusingwirelessnetworks}. It is particularly effective for dynamic environments, where reinforcement learning and real-time adjustments enhance accuracy, optimizing wireless indoor positioning systems for reliable localization even in variable conditions \cite{CarreraVillacres2019}.

The feature extraction capabilities of deep neural networks (DNNs) plays a critical role in enhancing the performance for indoor localization tasks. For instance, \cite{Kim2018} presented a scalable DNN architecture by processing the Wi-Fi signal data to classify the multi-building and multi-floor locations, thereby reducing the localization errors typically encountered in the traditional algorithms. Furthermore, long short-term memory (LSTM) - another type of DNN - is utilized for tracking the temporal dependencies within the received signal strength (RSS) data to better meet the demands of IoT-driven localization systems \cite{Ye2022}\cite{Hussain2019}.

Moreover, certain techniques employed in the extraction of channel state information (CSI) have emerged as a powerful tools for deep learning applications in indoor positioning. It has been shown that CSI can be transformed into time-frequency matrices that resemble images, enabling convolutional neural networks (CNNs) to perform spatial feature effectively \cite{Hsieh2019}. Deep learning frameworks indicate that employing CNNs alongside UWB signals enhances feature extraction capabilities typically in indoor environments \cite{Nosrati2024}. 

\section{Data Simulation}
\label{sec:data_simulation}

The development of deep learning models for indoor positioning systems requires comprehensive training datasets that accurately reflect the complexities of real-world environments. This section details the methodology used for simulating range and angle measurements, which form the foundation for constructing the likelihood grid maps used in training the ResNet model.

In previous works the challenges for radio propagation and radio-based localization associated with the aircraft cabin environment have been studied \cite{Schwarzbach_2020_Covid19, schmidtWirelessConnectivityAirplanes2021b, geyerPreciseOnboardAircraft2022}. To further elaborate on these challenges, we address operational simulation representing a boarding process in an A320 combined with a ray tracing and stochastical propagation and error modeling for radio propagation. The applied simulation framework for radio propagation builds upon two foundational works. Schwarzbach et al. \cite{Schwarzbach2022UWBStatisticalEvaluation} conducted an extensive survey of UWB ranging measurements in challenging environments, resulting in a statistical model. This model enables probabilistic generation of UWB range measurements with realistic characteristics, accounting for Line-of-Sight (LOS), Non-Line-of-Sight (NLOS), outlier reception scenarios, and measurement failures. The statistical distributions derived from their empirical data capture the error patterns typically encountered in real-world deployments. In addition, \cite{Schwarzbach2024SimulationIPSConnectedCabin} applied this statistical model in conjunction with ray tracing to determine anchor visibility within aircraft cabin environments. This approach accounts for the complex geometry and material properties of the aircraft, providing a comprehensive dataset published in \cite{Schwarzbach2023SimulationData}.

Building on these works, our simulation consists of several key components: 

\begin{itemize}
    \item \textbf{3D Modeling}: A detailed model of the aircraft cabin, including geometric dimensions and material properties.
    \item \textbf{Anchor Placement}: Multiple anchors are positioned throughout the cabin space based on optimal coverage analysis and practical installation constraints.
    \item \textbf{Ray Tracing:} Algorithms simulate signal propagation between anchors and potential tag positions, accounting propagation phenomena introduced by cabin structures.
    \item \textbf{Statistical error modeling:} The deterministic rays are augmented with statistical error models derived from \cite{Schwarzbach2022UWBStatisticalEvaluation}, introducing realistic measurement uncertainties.
\end{itemize}

\subsection{Overview of aircraft setup}
The aircraft cabin environment presents unique challenges for indoor positioning due to its elongated geometry, densely packed seating arrangement, and various obstructing elements. Our simulation recreates a standard single-aisle commercial aircraft cabin (dimensions: $30 \times 3.5 \times \SI{2.4}{\meter}$ representing an A320. The simulation incorporates $8$ anchor nodes spread throughout the plane. The configuration is visualized in \cref{fig:concept_diagram}. The reference input data for the localization system is derived from an operational simulation that models the aircraft boarding process. This simulation incorporates 148 passengers boarding the aircraft according to the methodological framework established by Schultz et al. \cite{schultz_fast_2018}. Each passenger is assumed to carry a wireless tag, constituting the mobile nodes to be localized within the aircraft cabin environment.


\subsection{Measurement simulation and error modeling}
For each passenger at each time step during the boarding process a potential tag position is derived. For this task, we use the reference positions of the boarding simulation. Based on the reference states, a true Euclidean distance $d_i$ and a true azimuth angle $\theta_i$ between the tag positions $\boldsymbol{\mathrm{x}}_i = (x_i, y_i)^{\intercal}$ and each anchor $\boldsymbol{\mathrm{x}}^a = (x^a, y^a)^{\intercal}$ is calculated with:

\begin{align}
    d_i^a &= \sqrt{(x_{i} - x^{a})^2 + (y_{i} - y^{a})^2} \label{equ:distance}\\
    \theta_{i}^a &= \arctan2\left(\frac{y_{i} - y^{a}}{x_{i} - x^{a}}\right) \label{equ:angle}\; .
\end{align}

The additive error components $\varepsilon_r$ for range measurements, are implemented based on the statistical error characterization framework established in \cite{Schwarzbach2022UWBStatisticalEvaluation}. Furthermore, the statistical properties of the AoA errors are derived from the work presented by Yu et al. in \cite{Yu2009AoAErrorModeling}. The measurement simulation follows a structured decision tree process: Beginning with the true relation (distance $d$, angle $\theta$), we first determine if a measurement failure occurred. In case of failure, a null value is recorded ($r = \text{None}$, $a = \text{None}$). For valid measurements, the simulation classifies the error type based on the ray tracing simulation into three categories: Line-of-Sight (LOS), Non-Line-of-Sight (NLOS), or outlier conditions. Each category then applies the corresponding stochastic model from \cref{tab:error_models}.

\begin{table*}[pos=h]
    \caption{Stochastic Error Models for ranging and angle measurements.}
	\label{tab:error_models}
	{\renewcommand{\arraystretch}{1.0}
		\begin{tabular}{p{1.25cm}p{3.5cm}p{3.5cm}}
        \toprule
        \textbf{Condition} & \textbf{Range} & \textbf{Angle} \\
        \midrule
        LOS & 
        $\varepsilon_r \leftarrow \mathcal{N}(0, \sigma_r^2)$ \newline $\sigma_r = 0.3$ m & 
        $\varepsilon_a \leftarrow \mathcal{N}(0, \sigma_{\theta}^2)$ \newline $\sigma_{\theta} = 3^{\circ}$ \\
        \hline
        NLOS & 
        $\varepsilon_r \leftarrow \mathcal{LN}(\mu, \sigma)$ \newline $\mu = 0.8$; $\sigma = 1.07$ & 
        $\varepsilon_a \leftarrow \mathcal{U}(-\pi, \pi)$ \\
        \hline
        Outlier & 
        $\varepsilon_r \leftarrow \mathcal{U}(-d, d)$ & 
        $\varepsilon_a \leftarrow \mathcal{U}(-\pi, \pi)$ \\
        \bottomrule
		\end{tabular}
	}
\end{table*}

\begin{figure}[pos=ht]
	\centering
	\begin{subfigure}[b]{0.48\textwidth}
		\centering
    	\includegraphics[trim=0 0 0 0, clip, width=1\linewidth]{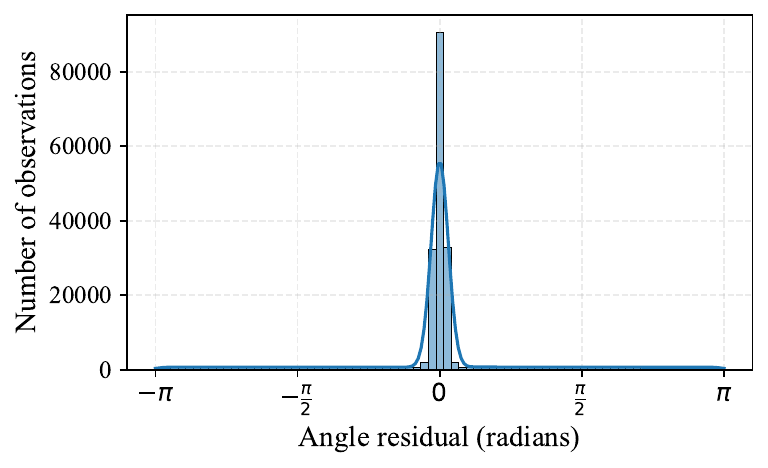}
    	\caption{}
    	\label{fig:histogram_angles}
	\end{subfigure}
	\hspace{0.2cm}
	\begin{subfigure}[b]{0.48\textwidth}
		\centering
    	\includegraphics[trim=0 0 0 0, clip, width=1\linewidth]{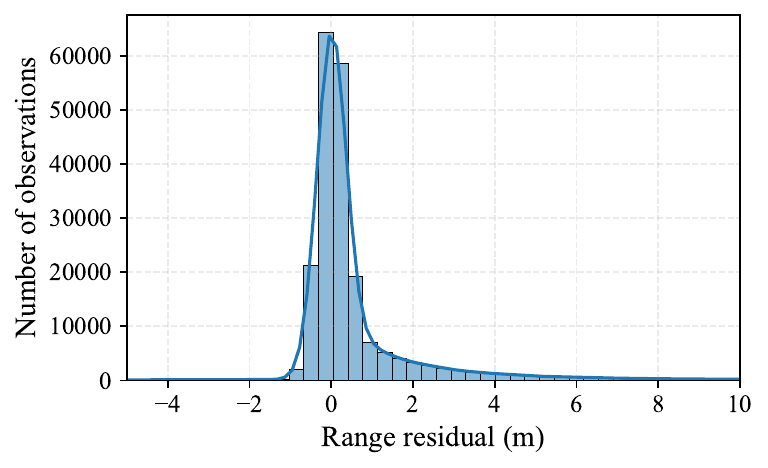}
    	\caption{}
    	\label{fig:histogram_ranges}
	\end{subfigure}%
	\centering
	\caption{Histograms showing the distribution of residuals for (a) angle measurements in radians and (b) range measurements in meters.}
	\label{fig:histograms}
\end{figure}
\section{Likelihood grid maps} 
\label{sec:grid_maps}


We implement a discrete state space representation utilizing the principles of grid-based Bayesian filtering. The spatial domain of interest is discretized into a uniform grid $G$ with dimensions $M \times N$, where each cell $\boldsymbol{\mathrm{g}}_{y(m,n)}$ represents a position hypothesis. The fundamental advantage of using these grids come from the fully convolutional nature of the ResNet model which has an innate ability to leverage spatial representations. It can train effectively on image grids allowing the model to capture localized features and exploit spatial patterns much better than the other DNN models due to the skip connections and the residual blocks feature \cite{Conneau2017}. 

\subsection{Known term calculation}
At first, for each grid cell $\boldsymbol{\mathrm{g}}_{(g,h)}$ and each anchor point $\boldsymbol{\mathrm{x}}^a$, we compute the known term values based on the given geometric relationships given in \cref{equ:distance} and \cref{equ:angle}. These values constitute the expected measurements under ideal conditions, derived directly from the geometric configuration.

\subsection{Residual calculation}

The measurement residuals are computed by comparing the known terms with the given observations ($\hat{d}$ and $\hat{\theta}$):

\begin{enumerate}
    \item \textbf{Range residual}: $r_{(g,h)} = d_{(g,h,k)} - \hat{d}_k$
    \item \textbf{Angle residual}: $r_{\theta,(g,h)} = \theta_{(g,h,k)} - \hat{\theta}_k$
\end{enumerate}

These residuals quantify the discrepancy between expected and observed measurements for each potential position, forming the basis for likelihood evaluation.

\subsection{Observation model}

The residuals are transformed into likelihood values through the application of an observation model that characterizes the probability of obtaining the observed measurement given a hypothesized position. While various formulations can be employed, a common approach utilizes a Gaussian observation model:

\begin{enumerate}
    \item \textbf{Range likelihood}: $L^{(r)}_{(g,h,k)} = \exp\left(-\frac{(r_{(g,h,k)})^2}{2\sigma^2_r}\right)$
    \item \textbf{Angle likelihood}: $L^{(\theta)}_{(g,h,k)} = \exp\left(-\frac{(r_{\theta,(g,h,k)})^2}{2\sigma^2_\theta}\right)$
\end{enumerate}

where $\sigma^2_r$ and $\sigma^2_\theta$ represent the variance parameters of the observation model for range and angle measurements, respectively. 

The resulting likelihood maps $L^{(r)}_{(g,h,k)}$ and $L^{(\theta)}_{(g,h,k)}$ provide a spatially-resolved representation of measurement compatibility, with higher values indicating greater probability of the tag being located at the corresponding position. For positions with multiple anchor measurements, the individual likelihood maps can be combined through multiplication (assuming conditional independence) to obtain an integrated likelihood field:

\begin{equation}
    L_{(g,h)} = \prod_{k=1}^{K} L^{(r)}_{(g,h,k)} \cdot L^{(\theta)}_{(g,h,k)}
\end{equation}

This integrated likelihood map serves as a comprehensive spatial representation of position probability based on the complete set of available measurements, forming the foundation for subsequent position estimation through maximum likelihood or Bayesian inference methods.  An exemplary output for the derived likelihood grid maps for both measurement types is given in the graphical abstract in \cref{fig:concept_diagram}.
\section{Positioning using DNNs}
\label{sec:positioning}

In this section, we explore how the deep neural networks has been implemented to estimate the location of the passengers in the simulated indoor cabin environments using the ResNet. The focus is on the model's architecture along with the inputs and loss function used for the model.

\begin{figure}[ht!]
\centerline{\includegraphics[width=0.6\linewidth]{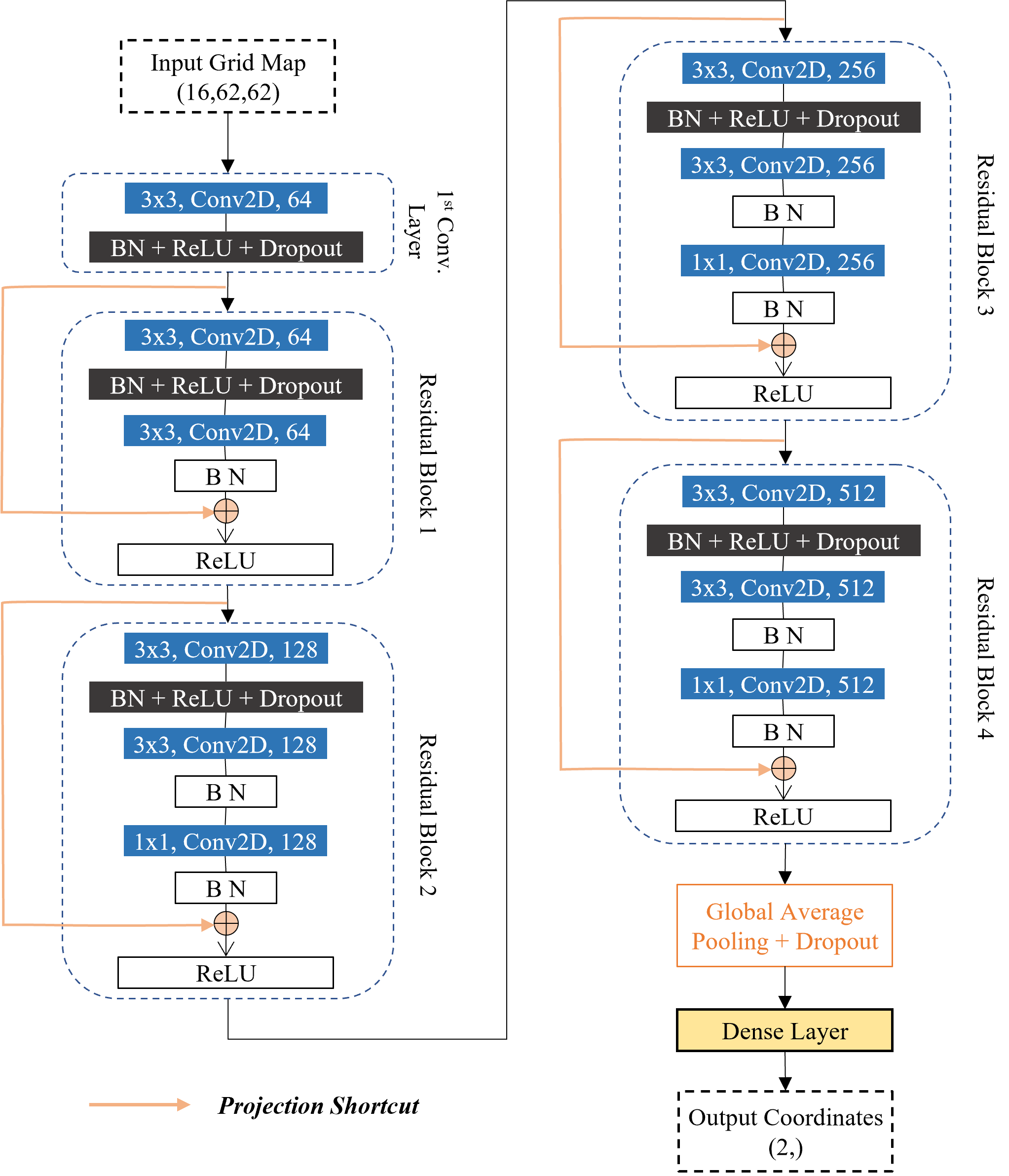}}
\caption{ResNet model architecture used for indoor localization}
\label{fig:resnet_architecture}
\end{figure}

\subsection{ResNet architecture}
ResNet model has been trained on the grid maps to predict the location coordinates. The deep layering inside the ResNet model allows it to learn the complex spatial features in the likelihood grid maps. 
The residual blocks in the ResNet model allows the input residual maps to train more effectively with better generalization despite having a complex and deeper architecture \cite{ResNetSolvingVanishingGradient}\cite{Niu2023}. The model's architecture shown in \cref{fig:resnet_architecture} consists of multiple residual blocks each including convolutional layers along with the batch normalization (BN) \cite{BatchNormalization}, rectified linear unit (ReLU) \cite{RELU}, and Dropout\cite{JMLR:v15:srivastava14a} layers. Furthermore, the addition of the dropout layers helps to mitigate overfitting making the model generalize better for unseen data \cite{JMLR:v15:srivastava14a}. 

Although the original ResNet model does not include a dropout layer, we incorporate it into our architecture to further stabilize the training, as also implemented in \cite{resnet_dropout}. The output of the convolutional layers inside the residual block is added back to the original input using a shortcut connection. This establishes an identity mapping \cite{identity_mapping} thus enhancing the gradient flow during back-propagation. To enable the model to predict continuous $x,y$ position coordinates, the output layer is implemented as a Dense layer with a linear activation function.

\subsection{Input and loss function}
The likelihood grid maps have been used as inputs for the ResNet model as shown in \cref{fig:concept_diagram}. For each measurement, we have the range and AoA grid maps for all the unique anchors but whenever there is a measurement failure, we pass an array of zeros with the same shape as our valid measurement grid. This issue is commonly present in real-world scenarios due to signal blockage or measurement limitations. Handling the invalid measurement is very important to ensure consistent input dimensions for the ResNet model.

Since both grid maps share the same dimensions, these are combined by stacking them along the channel dimensions. For \textit{\textbf{N}} anchor points, we have a total of $2 \times N$ channels. In our dataset, each measurement have ranges and angles from 8 anchors points leading to a total of 16 channels - 8 channels for each measurement's likelihood maps. Consequently, the input array has a dimension of $16 \times 16 \times 62$ as illustrated in \cref{fig:resnet_architecture}, where $62 \times 62$ represents the spatial grid shape.

During the training process, the model must also get the ground truth labels since ResNet is a supervised learning algorithm. The ground truth labels consists of the true target locations represented as a $(x, y)$ coordinates in 2D plane. For each training epoch, the grid maps are paired with the corresponding true locations, allowing the model to learn and minimize the error between the predicted and actual locations. 

The mean squared error (MSE) is a regression loss function which is used for our model to minimize the error. This function guides the model optimizing towards getting accurate position predictions. It calculates the average squared difference between predicted and actual coordinates \cite{MSE}. The function is mathematically expressed as 
\begin{equation}
MSE = \frac{1}{n} \sum_{i=1}^{n} (y\_pred - y\_true)^2,
\end{equation}

where \textit{y\_pred} represents the $(x, y)$ coordinates predicted by the model and \textit{y\_true} represents the actual $(x, y)$ coordinates.

\section{Performance evaluation}
\label{sec:evaluation}

To evaluate the performance of any deep learning model, it is crucial to ascertain the optimal hyperparameter settings that would enhance the model performance significantly. Hyperparameter tuning is carried out to avoid overfitting as well as optimizing the performance and generalization of machine learning models \cite{AIlemobayo2024}. The choice of hyperparameters significantly influences the efficiency of training and accuracy of predictions. Before training the model, we implemented a systematic hyperparameter search to identify the best configurations and then train the ResNet model on those settings.

\subsection{Impact of hyperparameter selection on accuracy}
The hyperparameters selected include learning rate, dropout rate, batch size, and optimizer, as summarized in \cref{tab:hyperparameters_choice}. The hyperparameter space from which these were drawn is also included in the same table, alongside the best hyperparameter settings concluded from trials based on performance outcomes shown in \cref{tab:hyperparameters_trials}.

\begin{table}[th!]
\centering
\renewcommand{\arraystretch}{1.1}
\caption{Hyperparameter search space and selected hyperparameters for the ResNet model}
\label{tab:hyperparameters_choice}
\begin{tabular}{@{}lp{4cm}l@{}}
\toprule
    \textbf{Hyperparameter}    &  \textbf{Search Space} & \textbf{Selected Values}                          \\ \midrule
    Optimizer          & ['Adam'\cite{adam_adamax}, 'Adamax'\cite{adam_adamax}, 'Adagrad'\cite{adagrad_ref}, 'RMSprop'\cite{rmsprop}] & \texttt{Adam}         \\ 
    Learning Rate          & [0.01, 0.001, 0.0005, 0.0001] & \texttt{0.001}         \\ 
    Batch Size           & [8, 16, 32, 64] & \texttt{32} \\ 
    Dropout Rate           & [0.2, 0.3, 0.4, 0.5]  & \texttt{0.2}         \\ 
    \bottomrule
\end{tabular}
\end{table}

To identify the optimal set of hyperparameters for our model, we employed a Bayesian Optimization \cite{BayesianOptimization}, a statistical technique applied through Keras \cite{chollet2015keras} library to streamline the search for effective hyperparameter settings. We have applied 12 iterations on different combinations of hyperparameters and evaluated the loss on validation data to track the model's performance. This metric termed as validation loss $(val\_loss)$ is in meters and it serves as an indicator of the model's capacity to generalize on unseen data.

\begin{table}[ht!] 
    \centering
    \renewcommand{\arraystretch}{1.1}
    \setlength{\tabcolsep}{1.5pt} 
    \caption{Hyperparameter search trials based on Bayesian Optimization with the validation loss in meters}
    \begin{tabular}{c c c c c c}
        \toprule
        \textbf{Trial No.} & \textbf{Optimizer} & \textbf{Learning Rate} & \textbf{Batch Size} & \textbf{Dropout Rate} & \textbf{Val\_loss} \\
        \midrule
        1  & Adagrad & 0.0001 & 32  & 0.2 & 3.724  \\
        2  & Adagrad & 0.0005 & 8   & 0.3 & 15.149 \\
        3  & Adagrad & 0.0001 & 8   & 0.5 & 26.267 \\
        4  & RMSProp & 0.0005 & 16  & 0.4 & 0.105  \\
        5  & Adamax  & 0.0001 & 64  & 0.3 & 0.448  \\
        6  & RMSProp & 0.01   & 8   & 0.2 & 0.096  \\
        7  & Adam    & 0.0001 & 16  & 0.3 & 0.116  \\
        8  & Adam    & 0.01   & 16  & 0.5 & 0.153  \\
        9  & Adagrad & 0.01   & 64  & 0.5 & 0.329  \\
        \textbf{10} & \textbf{Adam}    & \textbf{0.001}  & \textbf{32}  & \textbf{0.2} & \textbf{0.084}  \\
        11 & Adam    & 0.001  & 8   & 0.3 & 0.095  \\
        12 & Adamax  & 0.001  & 16  & 0.3 & 0.106  \\
        \bottomrule
    \end{tabular}
    \label{tab:hyperparameters_trials}
\end{table}

As shown in \cref{tab:hyperparameters_trials}, the model reaches its optimal performance in trial number 10 with the least $val\_loss$, utilizing Adam as the optimizer, a learning rate of 0.001, a batch size of 32, and a dropout rate of 0.2. Using Adam an optimizer has been favorable in many deep learning applications since it adaptively adjusts the learning rate, facilitating more efficient convergence rates compared to traditional optimizers \cite{Zhou2024}. Additionally, trial 1 to 3 show that the model performance is declined significantly with the utilization of Adagrad optimizer with smaller batch sizes.

\subsection{Results}

\begin{figure}[pos=ht]
	\centering
	\begin{subfigure}[b]{0.48\textwidth}
		\centering
    	\includegraphics[trim=0 0 0 0, clip, width=1\linewidth]{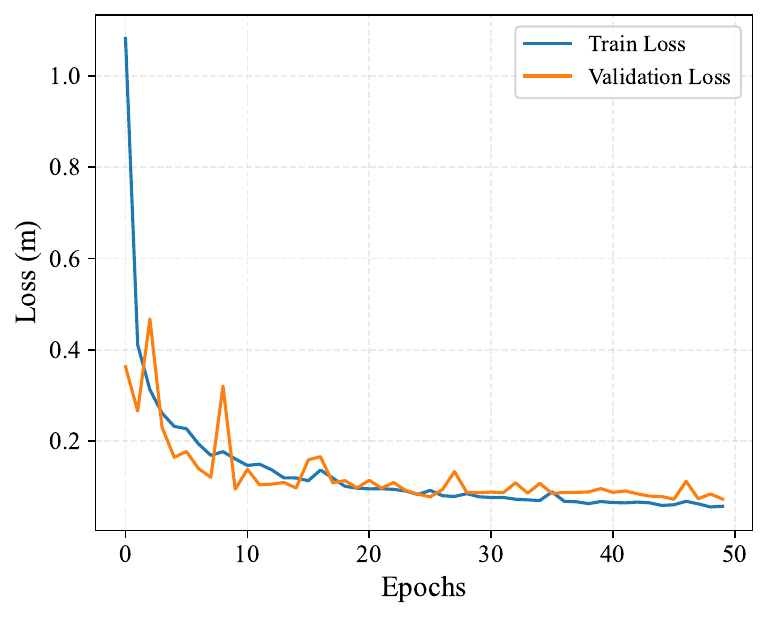}
    	\caption{}
    	\label{fig:loss_over_epochs}
	\end{subfigure}
	\hspace{0.2cm}
	\begin{subfigure}[b]{0.48\textwidth}
		\centering
    	\includegraphics[trim=0 0 0 0, clip, width=1.01\linewidth]{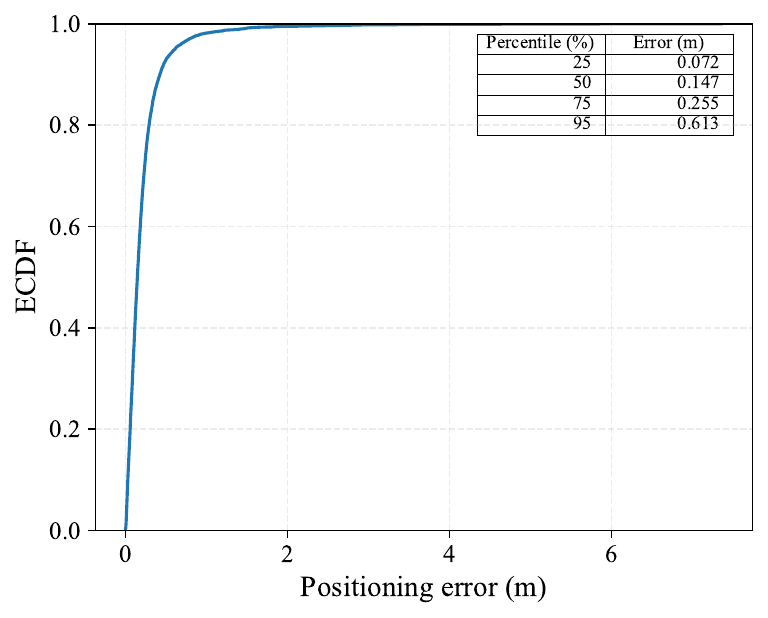}
    	\caption{}
    	\label{fig:ecdf_plot}
	\end{subfigure}%
	\centering
	\caption{(a) Training and validation loss curves over 50 training epochs. (b) ECDF of positioning error with key percentiles in meters.}
	\label{fig:results}
\end{figure}

The output of our ResNet model provides the $x, y$ coordinates of the predicted location based on the input grid maps. An example output of predicted position is shown in graphical abstract in \cref{fig:concept_diagram}. The model's performance has been assessed using test dataset comprising $25\%$ of the total dataset, while the remaining $75\%$ is allocated for training the ResNet model. The model is trained over 50 epochs, as illustrated in \cref{fig:loss_over_epochs}, which also displays both the training and validation loss (in meters) throughout the epochs. It is observed that the loss significantly decreases as the epoch count increases; however, it begins to flatten out after approximately 30\textsuperscript{th} epoch.

During the training process, we defined a custom checkpoint using Keras functionality to save the trained model if the accuracy of the subsequent epoch improves. The last saved model with the best accuracy came out to be at 46\textsuperscript{th} epoch with the validation loss equal to $\SI{7.28}{\centi\meter}$ on test dataset. This accuracy is consistent with, and in many cases exceeds, existing positioning accuracies for indoor localization that employs technologies such as Wi-Fi, UWB, or BLE for positioning.

\cref{fig:ecdf_plot} illustrates the positioning errors measured in meters as empirical cumulative distribution function (ECDF) based on the test data. We observe that the majority of the positioning errors are concentrated within a small range, indicating high positioning accuracy for majority of the observations. The analysis shows a positioning error of $\SI{0.147}{\meter}$ for the 50\textsuperscript{th} percentile which means that $50\%$ of the predicted locations have an error within the range of $\SI{14.7}{\centi\meter}$. Beyond $\SI{0.613}{\meter}$ (95\textsuperscript{th} percentile), the curve plateaus, meaning that only a small fraction of observations exhibit higher errors. Although fewer in number, the slight extension of the tail indicates the presence of outliers.


\section{Conclusion}
\label{sec:conclusion}

Recent advancements in artificial intelligence have established a new potential in indoor localization by providing a reliable and accurate positioning in indoor environments. In this study, we train ResNet model to learn the underlying spatial patterns in the likelihood grid maps generated based on the simulated range and AoA measurements. The likelihood grid maps, which are used as model inputs, represent the pixel-level position hypothesis. Before training the model, hyperparameter tuning is performed to get the best training parameters since DNNs always involve multiple parameter settings which can tweak the positioning performance. The attainment of a model's accuracy of $\SI{7.28}{\centi\meter}$ signifies a potential for advanced positioning in indoor environments. The trained model is then used to predict positions on test dataset and the results showed that $95\%$ of the predictions have a positioning error less than $\SI{0.613}{\meter}$, with a median error of only $\SI{0.147}{\meter}$. This level of precision underscores the effectiveness of our approach and lays a robust groundwork for future enhancement in indoor localization systems.

\begin{acknowledgments}

This work has been funded by the German Federal Ministry for Economic Affairs and Climate Action (BMWK) within the project INTACT (FKZ: 20D2128D).

\begin{figure}[pos=h]
\centering
\includegraphics[width=0.25\textwidth]{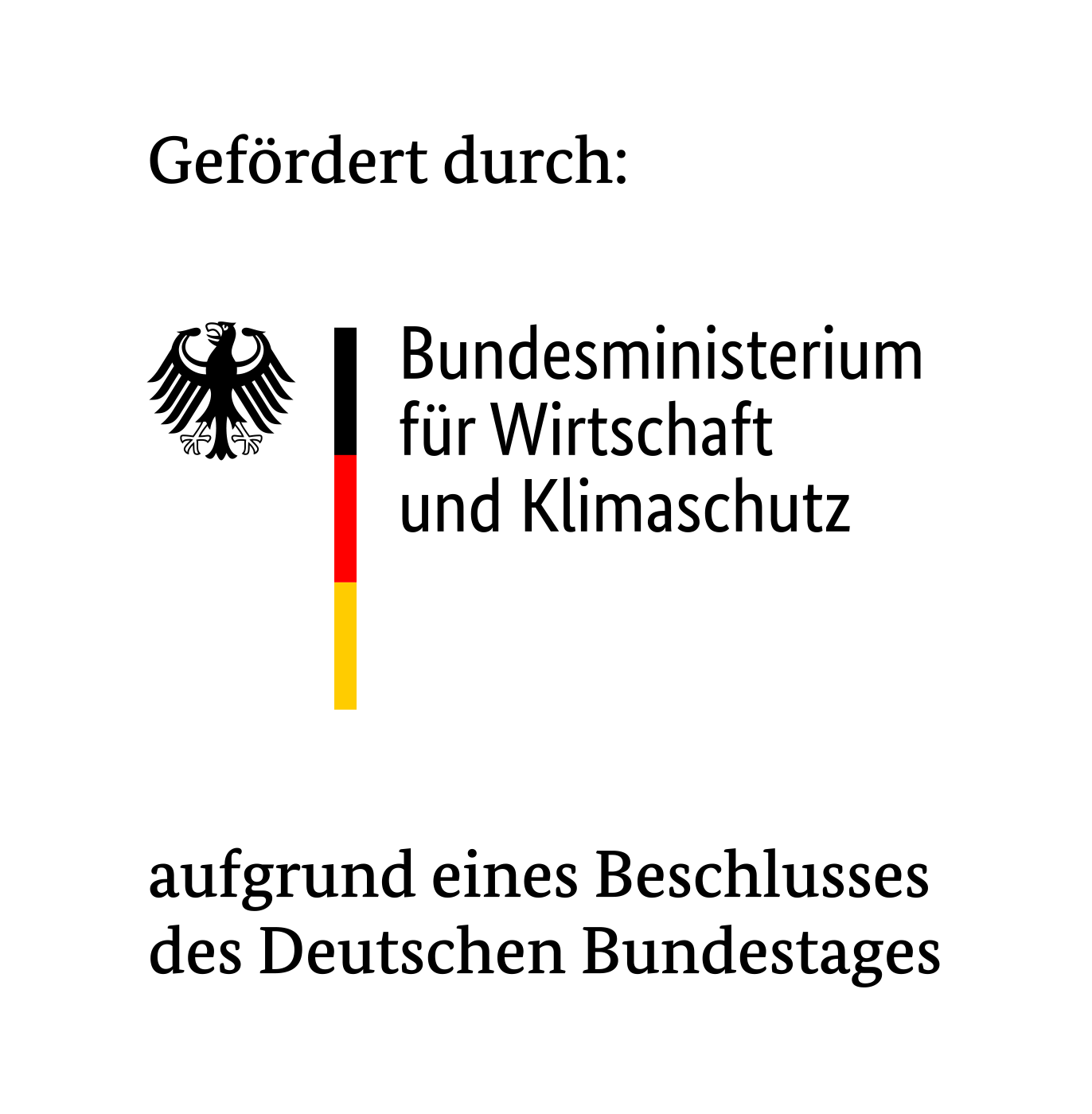}
\end{figure}
\vspace{-0.5cm}

\end{acknowledgments}

\section*{Declaration on Generative AI}
The author(s) have not employed any Generative AI tools.
  
\bibliography{main}

\end{document}